\documentclass[%
 reprint,
superscriptaddress,
 showpacs,preprintnumbers,
 bibnotes,
 amsmath,amssymb,
 aps,
 prb,
]{revtex4-1}

\usepackage{graphicx}
\usepackage{dcolumn}
\usepackage{bm}
\usepackage{epstopdf}

\usepackage[utf8]{inputenc}
\usepackage{color}
\usepackage{float}
\usepackage{amsfonts}

\begin{document}


\title{Generation of spin-dependent coherent states in a quantum wire}
\author{J. Paw\l{}owski}
\affiliation{
	Faculty of Physics and Applied Computer Science,
	AGH University of Science and Technology, Krak\'{o}w, Poland}
\author{P. Szumniak}
\affiliation{Department of Physics, University of Basel, Klingelbergstrasse 82, 4056 Basel, Switzerland}
\affiliation{
	Faculty of Physics and Applied Computer Science,
	AGH University of Science and Technology, Krak\'{o}w, Poland}
\author{S. Bednarek}
\affiliation{
	Faculty of Physics and Applied Computer Science,
	AGH University of Science and Technology, Krak\'{o}w, Poland}

\date{\today}

\begin{abstract}
We propose an all-electrically controlled nanodevice---a gated semiconductor nanowire---capable of generating a coherent state of a single electron trapped in a harmonic oscillator or superposition of such coherent states---the Schr\"odinger cat state.
In the proposed scheme, electron in the ground state of the harmonic potential is driven by resonantly oscillating Rashba spin-orbit coupling.
This allows for the creation of the Schr\"odinger cat state with superposition amplitudes depending on the initial electron spin state. Such a method can be used for initialization of a single spin qubit defined in a coherent state.
The harmonic confinement potential along the InSb nanowire and the modulation of the Rashba spin-orbit coupling is obtained by proper gating. 
The results are supported by realistic three-dimensional time-dependent self consistent Poisson-Schr\"odinger calculations.


\end{abstract}


\maketitle


\textit{Introduction}.
There is currently great interest in studying spin related phenomena in semiconductor quantum nanowires with spin-orbit interaction (SOI) \cite{16,16a,16b}. Such spin-orbit nanowires are particularly promising for the realization of spin based quantum information processing\cite{PhysRevA.57.120, *CKRevl} and the generation of exotic quasi-particles\cite{PhysRevLett.100.096407,PhysRevLett.105.077001,24,25,26,111} in solid state systems.
The Rashba type SOI (RSOI) \cite{BR1,*BR2,4} originates from the asymmetry of the confining potential and can be tuned with external gate voltages.
Proper gating of semiconductor nanostructures allows for the realization of dynamic and spatial modulation of RSOI. This may lead to many interesting effects, which have recently been studied in different systems \cite{27,28,29,30,31,32,33,34,35,36,1,47}.

Coherent states (CSs) of a harmonic oscillator \cite{19} may be used in optical quantum computing \cite{61,62,63,64,65,66,67,68,69}, utilizing trapped ions \cite{57,58,59}. In one of several approaches, a qubit is defined in the basis of two CSs with opposite amplitudes\cite{21,22,23,2,3,40}. These two CSs with opposite displacements should be well-separated, forming, in superposition, the so-called Schr\"odinger cat state \cite{71,72,73,74,75,76,3,2,39}. In this unusual approach, the computational basis states are only approximately orthogonal \cite{23,2}.

In [\onlinecite{7,10}], the authors show a scheme for performing quantum logic operations in solid-state systems exploiting spins of a single electron in the CS, which forms a so-called coherent-state spin qubit. 
Interestingly, one can generate a maximally entangled state between such qubits, defined as CSs localized on either side of a 1D harmonic oscillator \cite{7,10}, or generally on local spins defined on two sides of a quantum dot \cite{32,81}. Here, the qubit is ``localized'' in space, but it can also be defined as the presence of an electron on each side of a double-dot (or wire) structure \cite{20,37,38,41}---a so-called charge qubit. 

In the current paper, we exploit the effects of time-dependent RSOI to propose a method for preparing a single electron in superposition of two CSs with opposite displacements. Such superposition amplitudes depend on the initial spin state of the electron. If the spin is aligned parallel to the $z$-axis, we get a single CS which
leads us to prepare a coherent-state spin qubit with the required amplitude. 
In contrast, for a spin perpendicular to the $z$-axis, we get an equal superposition of the CSs---Schr\"odinger cat state. This may be seen as a simple solid-state analog to the famous experiments producing the Schr\"odinger cat state of a trapped ion by D.~J.~Wineland and others \cite{71,72,73,74,75,76}.
It is worth mentioning that there are 
other methods suitable for preparing CSs, which exploit surface acoustic waves\cite{8,9}, or the conditional displacement mechanism\cite{82}.


We propose a scheme which can be implemented in a gated semiconductor nanowire. We provide a detailed realistic model of such structures\cite{47}. The proper gating allows the generation of the appropriate harmonic potential and enables the induction of RSO coupling, modulated by a lateral electric field. 
We show that the lateral field does not destroy the generated CSs.
The proposed method, combined with a charge sensing scheme \cite{45,38,46}, may also serve as a magnetic field-free electron spin readout.


\textit{Device model and calculation method}.
\begin{figure}[b,t]
\includegraphics[width=7.5cm]{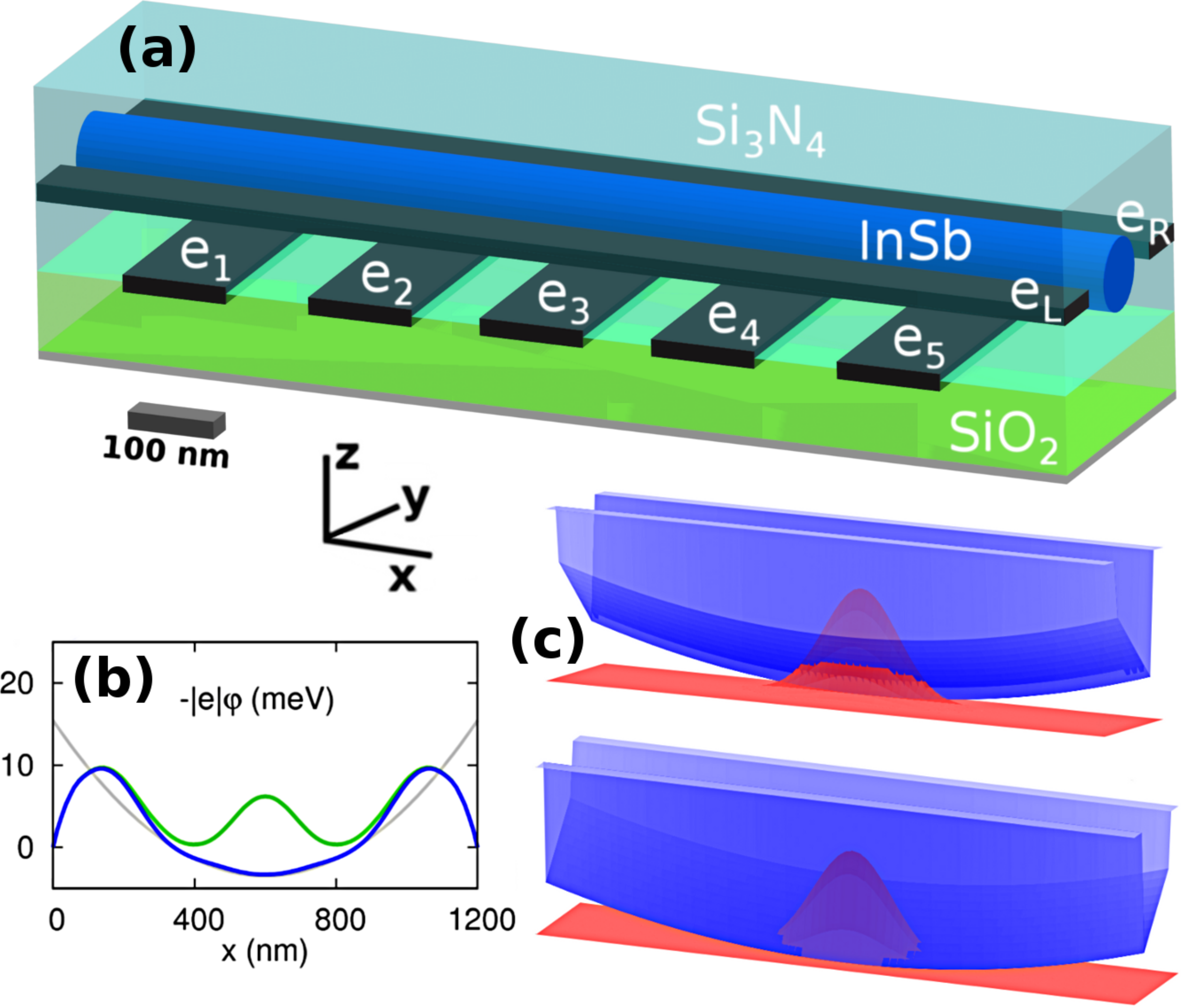}
\caption{\label{fig:1} The schematic view of the nanodevice. (a) The presented structure consists of a gated nanowire. (b) A five bottom gates form a parabolic-shaped quantum dot (blue) along the wire\cite{52}. (c) Two side gates create a lateral (in the $y$ direction) electric field, which is the source of the oscillating RSOI, which modifies the parabolic confinement.}
\end{figure}
The proposed nanodevice is composed of a gated nanowire, as depicted in Fig.~\ref{fig:1}(a), where the quantum dot is defined by local gating in a similar manner as in the experimental set-ups~\cite{16,17,18}. The entire structure is placed on a strongly doped silicon substrate, covered with a 100~nm layer of $\mathrm{SiO}_2$. 
An array of five $120$~nm wide metallic gates ($\mathrm{e}_1$-$\mathrm{e}_5$) is deposited on the substrate. The metallic gates are further covered with a 200~nm layer of $\mathrm{Si}_3\mathrm{N}_4$. An InSb nanowire with a diameter of 60~nm is deposited within (in the middle) the $\mathrm{Si}_3\mathrm{N}_4$ layer. Two side gates ($\mathrm{e}_L$,$\mathrm{e}_R$) are added on both sides of the wire. The $\mathrm{Si}_3\mathrm{N}_4$ layer isolates the gates from the nanowire.

The Hamiltonian for a single electron of effective mass $m$, confined within the wire described by the potential $\varphi(\mathbf{r},t)$, may be written in the following form:
\begin{equation}
H(\mathbf{r},t)=\left(-\frac{\hbar ^2}{2m}\boldsymbol\nabla^2-|e|\varphi(\mathbf{r},t)\right)1_2+H_R(\mathbf{r},t),
\label{ham}
\end{equation}
with InSb band mass $m=0.014 m_e$. The space-dependent electric field introduces the inhomogeneous RSOI\cite{101}:
$H_R(\mathbf{r},t)=\gamma_\mathrm{3D}e(\mathbf{E}(\mathbf{r},t)\times \mathbf{k})\cdot\boldsymbol\sigma$,
with the electric field $\mathbf{E}(\mathbf{r},t)=-\boldsymbol\nabla \varphi(\mathbf{r},t)$, the wave vector  $\mathbf{k}=-i\boldsymbol\nabla $, $\boldsymbol\nabla \equiv \left[\partial _x,\partial _y,\partial _z\right]$, and the Pauli vector: $\boldsymbol\sigma \equiv \left[\sigma _x,\sigma _y,\sigma _z\right]$.

This implementation includes a relatively wide wire together with surrounding gates, isolating layers and substrate.
Thus, it is necessary to find the 3D confinement potential $\varphi(\mathbf{r},t)$ by solving the Poisson equation for the considered realistic structure.
We apply the calculation method described in Ref.~[\onlinecite{47}].
In our scheme, the control voltages applied to the side gates oscillate over time. Thus, in order to take into account the time dependence of the $\varphi (\mathbf{r},t)$ we solve the Poisson equation self consistently with a time-dependent Schr\"odinger equation:  
\begin{equation}
i\hbar\frac{\partial}{\partial t}\Psi(\mathbf{r},t)= H(\mathbf{r},t) \Psi(\mathbf{r},t).
\label{schr}
\end{equation}
The spin-1/2 electron wave function has a two-row spinor form:  $\Psi(\mathbf{r},t)=\left(\psi_{\downarrow }(\mathbf{r},t),\psi _{\uparrow }(\mathbf{r},t)\right)^T$, where the arrow indicates the spin projection onto the quantization axis ($z$).



Gates $\mathrm{e}_1$ and $\mathrm{e}_5$ form tunnel barriers at both ends of the wire and thus create an elongated quantum dot region within the wire.
The confinement $-|e|\varphi(\mathbf{r},t)$ along the wire\cite{52} depicted in Fig.~\ref{fig:1}(b)(blue curve) has a parabolic-like shape generated by the three middle gates $\mathrm{e}_{2,3,4}$. 
The voltages applied to these gates are: $V_1=V_5=-30$~mV, $V_2=V_4=10$~mV, $V_3=20$~mV.
The voltages applied to the two side gates $\mathrm{e}_L$ and $\mathrm{e}_R$ generate a lateral electric field which modifies the confinement potential in the $y$-direction (see Fig.~\ref{fig:1}(c)).
This lateral field is the main source of the oscillating RSOI.
Oscillating voltages are applied to the side gates: $V_L(t)=+V_\mathrm{lr}\cos(\omega t)$ and $V_R(t)=-V_\mathrm{lr}\cos(\omega t)$, with $V_\mathrm{lr}=500$~mV.

\textit{Analytic results for simplified 1D model}.
In order to illustrate the basic concept of the proposed scheme and understand the effect of the time dependent RSOI on the ground state of the harmonic oscillator, we consider an effective 1D problem---for a while we freeze motion in directions perpendicular to the wire: $y$, $z$. In the later part, where we discuss a particular proposal for the realization of the nanodevice, we will return to full 3D calculations.
Let us now consider a one-dimensional form of the Hamiltonian (\ref{ham}) for the single electron trapped in the wire placed parallel to the $x$-axis:
\begin{equation}
H_\mathrm{1d}(x)=\left(-\frac{\hbar ^2}{2m}\partial _x^2+u(x)\right)1_2+h_R,
\label{ham1d}
\end{equation}
where $u(x)$ is the confinement potential along the wire, and the last term accounts for effective 1D RSOI. We assume that the electric field has a component in the $y$ direction (perpendicular to the wire direction). Thus, the 1D RSOI Hamiltonian takes the following form: $h_R=-\gamma_\mathrm{3D}eE_y\sigma_z k_x=i\gamma_\mathrm{3D}eE_y\sigma_z\partial_x$.

If the confinement potential along the wire has a parabolic shape $u(x)=m\omega^2x^2/2$, we can solve the eigenproblem (\ref{ham1d}) analytically in the momentum space, and then transform the results back to the position representation\cite{103}. The resulting ground state is doubly degenerate due to the spin
$\Psi_{\pm}(x)\sim\exp{\left(-\beta x^2\right)}\exp{\left(\pm iqx \right)}\chi_{z\pm}$, where $\chi_{z+}\!\!=$\footnotesize{$\left(\begin{matrix}1\\0\end{matrix}\right)$}\normalsize, $\chi_{z-}\!\!=$\footnotesize{$\left(\begin{matrix}0\\1\end{matrix}\right)$}\normalsize, and $q = m\gamma_\mathrm{3D}e E_y/\hbar^2$. This well known result shows that SOI adds displacement to harmonic potential wave functions in the momentum-space.\cite{13,14,15}

The corresponding ground state wave functions in the position representation form Gaussians multiplied by plane waves with positive or negative wave numbers for different components of the spinor.
Note that the value of the wave vector $q$ does not depend on the curvature (frequency $\omega/2\pi$) of the harmonic potential and is proportional to the RSOI coupling. RSOI introduces the energy correction: $\mathit{\Delta E}=\frac{\hbar^2q^2}{2m}$.\cite{12}

If the initial electron spin state includes both non-zero spinor components, then after a sudden turn off of the SOI, each component will gain opposite momentum $\hbar q$ or $-\hbar q$.
In this way, one can obtain a spatial separation of the spin-density.
A similar effect is achieved when we suddenly turn on the SOI, but then $q$ becomes $-q$ and vice versa.

The same result can be obtained in a simpler way by using the ladder operators formalism.
The 1D Hamiltonian (\ref{ham1d}) in lowering
$a=\sqrt{\frac{m\omega}{2\hbar}}x+\sqrt{\frac{\hbar}{2 m\omega}}\partial_x$
and raising $a^\dag$ operators
representation is
$\left(a^\dag{}a+\frac 1 2\right)\hbar\omega +i\gamma_\mathrm{3D}eE_y\sigma_z\sqrt{\frac{m\omega}{2\hbar}}(a-a^\dag)$.
Let us note that the canonical transformation of the ladder operators:
$\tilde{a}=a-i\xi$, where $\xi =\frac{\gamma_\mathrm{3D}eE_y}{\hbar}\sigma_z\sqrt{\frac{m}{2\hbar\omega}}$ removes the RSOI term.
To get displacement $i\xi$ of the ladder operators, we need to make a similarity transformation $D(i\xi){}a{}D^\dag(i\xi)=a-i\xi$ with the help of the displacement operator 
$D(\alpha )=\exp(\alpha a^\dag-\alpha^\ast a)$ with displacement parameter $\alpha$. 
So, once again we get the result that RSOI makes the unperturbed state $\Psi$ displaced.
Explicitly, after inserting $\xi=\frac{q\hbar}{m}\sigma_z\sqrt{\frac{m}{2\hbar\omega}}$ we get  
\begin{equation*}
D(i\xi)\Psi=\left(\begin{matrix}e^{iqx}&0\\0&e^{-iqx}\end{matrix}\right)\Psi.
\end{equation*}


In the next step, we analyze the effect of a smooth change of the RSOI, in particular an oscillatory-like change.
Now, the 1D Hamiltonian (\ref{ham1d}) is time dependent $H_\mathrm{1d}(x,t)$
with an oscillating $h_R(t)=i\gamma_\mathrm{3D}eE_{y0}\cos(\omega t)\sigma_z\partial_x$ RSOI part, and an oscillation frequency $\omega/2\pi$ equal to the oscillator resonant frequency.
Simple but tedious calculations of the Magnus expansion \cite{42,43,44} for the time evolution operator for the Hamiltonian $H_\mathrm{1d}(x,t)$ up to the 3-rd term give, i.a., the expression
\begin{equation*}
e^{(\alpha(t) a^\dag-\alpha^\ast(t) a)\sigma_z}=D\left(\alpha(t)\sigma_z\right)=\left(\begin{matrix}D(\alpha(t))&0\\0&D(-\alpha(t))\end{matrix}\right).
\end{equation*}
Again, the displacement operator appears, but now with the time-dependent displacement parameter $\alpha(t)=e^{i\omega t}\sqrt{\frac{m\omega}{2\hbar }}\frac{\gamma_\mathrm{3D}eE_{y0}}{2\hbar} t$.
Oscillating SOI makes an initial state displaced, with the displacement parameter linearly increasing over time and opposite signs for different spinor components.
These results are in analogous to the classical harmonic oscillator with resonant driving where amplitude linearly increases over time. If the initial state is the oscillator ground state, during the evolution we obtain a CS or a combination thereof, depending on the spin.
This is because the displacement operator $D(\alpha)$ generates a CS $|\alpha\rangle$ from the ground state $|0\rangle$. 
As a result, each spinor component will be a CS with opposite displacement.
In particular, for the spin aligned in the $x$ direction we can get equal superposition of these two CSs, $|\alpha\rangle+|\!-\!\alpha\rangle$, i.e. Schr\"odinger cat state.
This gives an idea of spin-dependent CSs generation.

\textit{Simulation results and discussion}.
Now we present the results of the numerical simulation of a realistic device, which can generate spin-dependent CSs.
First, we introduce a simple method for testing if states obtained in the simulations are the CSs, and also to verify the analytical results presented above.
The lowering operator $a$ is non-hermitian; hence, its variance is defined as \cite{5}:
$\left(\mathit{\Delta a}\right)^2=\langle a^\dag{}a\rangle -\langle a^\dag\rangle \langle a\rangle$. 
This follows the uncertainty relation \cite{6}:
$\langle a^\dag{}a\rangle -\langle a^\dag\rangle \langle a\rangle \ge 0$,
where we have equality for eigenstates of the $a$ operator.
The CSs  $|\alpha \rangle$ are the $a$ eigenstates with eigenvalues  $\alpha $, i.e.
$a|\alpha\rangle=\alpha|\alpha\rangle$.
This suggests a method for checking if a given state is coherent. We can simply verify if quantity $c_{\uparrow}=\langle \psi _{\uparrow }|a^\dag{}a|\psi _{\uparrow }\rangle /|\langle \psi _{\uparrow }|a|\psi_{\uparrow }\rangle|^2$ equals one. The expectation values are calculated for the actual spinor components. 
If both spinor components forms CSs, the eigenvalues parametrize our CSs:  $\alpha _{\uparrow }=\langle \psi _{\uparrow }|a|\psi _{\uparrow }\rangle /\langle \psi _{\uparrow }|\psi _{\uparrow }\rangle $, where we add normalization.
There is a similar consideration for the second spinor component $\psi_\downarrow$.

In our simulations, a single electron is initially trapped in the wire, in the ground state $\psi_0(x)$ of the harmonic confinement potential, as presented in Fig.~\ref{fig:1}(b).
We investigated two different limit cases. In the first case the electron spin is pointing along the $x$ axis, which means that the initial state is of the form $(e^{iq_0 x},e^{-iq_0 x})^T\psi_0/\sqrt{2}$,
with the initial coupling value $q_0=m\gamma_\mathrm{3D}e E_{y0}/\hbar^2$.
In the second case the spin is aligned along the $z$ axis and the corresponding initial state has the form: $(e^{iq_0x},0)^T\psi_0$. \footnote{In our calculations this initial states were additionally relaxed to real ground states in full 3D wire structure by the imaginary time evolution method\cite{112}.}

\begin{figure}[t]
\includegraphics[angle=0,width=8.7cm]{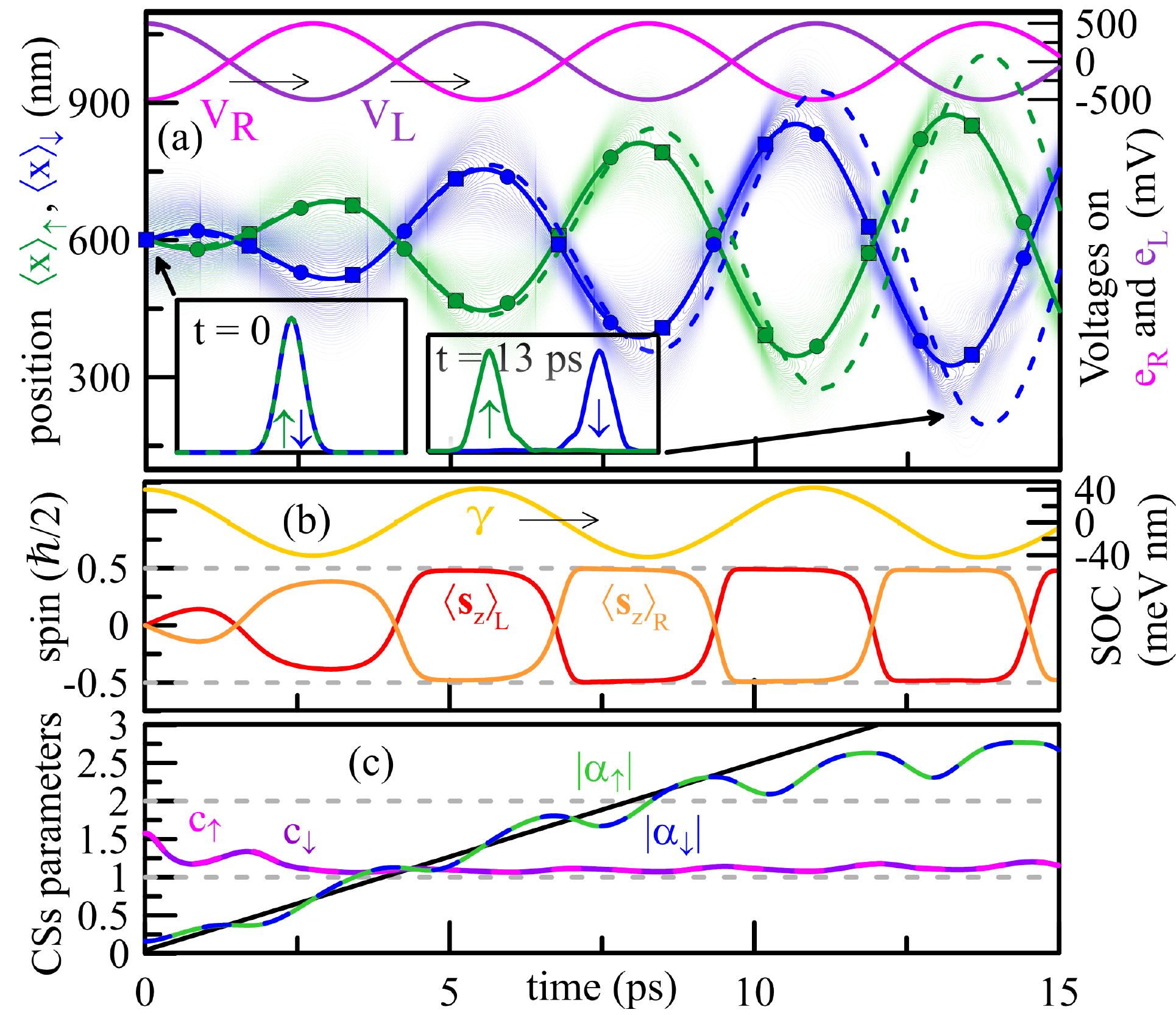}
\caption{\label{fig:2} (a) The variable side voltages (purple and magenta) generate oscillations of the spinor components (green $|\psi_\uparrow|^2$ and blue $|\psi_\downarrow|^2$), (b) mediated by the oscillating RSOI (yellow). (c) Both components form CSs  with linearly increasing amplitudes (green and blue).}
\end{figure}

\begin{figure}[t]
\includegraphics[angle=0,width=8.7cm]{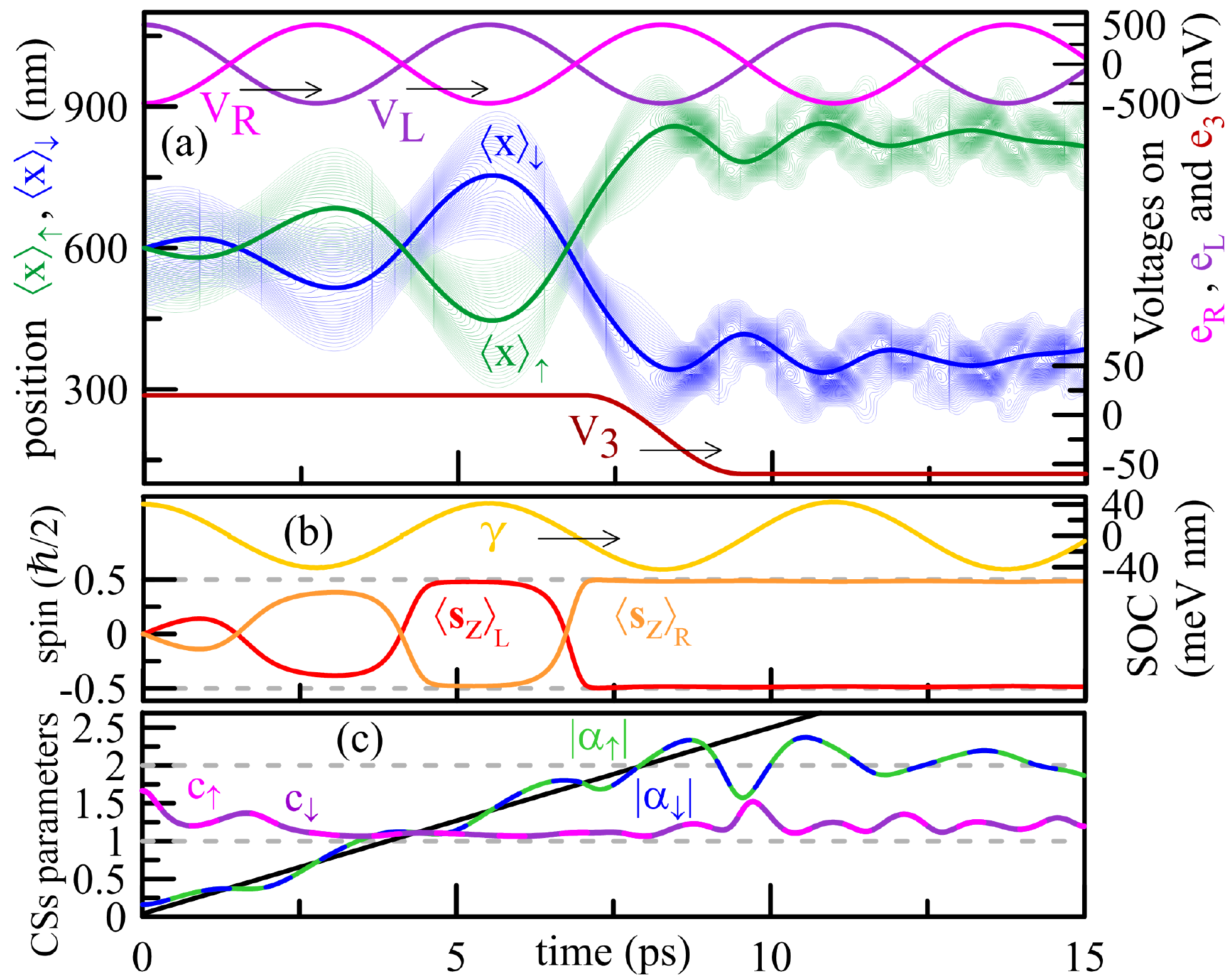}
\caption{\label{fig:3} The same generation process as in Fig.~\ref{fig:2}, but with the additional CS stabilization which starts at $t=7$~ps. Here, we see that the modules $|\alpha_{\uparrow,\downarrow}|$ stabilize around 2.}
\end{figure}

Let us now examine the nanodevice evolution, governed by Eq.~(\ref{schr}).
We turn on the oscillations of the side gates voltages $V_L(t)$ and $V_R(t)$, denoted by purple and magenta curves in Fig.~\ref{fig:2}(a). 
The voltages oscillation frequency $\omega/2\pi$ (represented by the gray curve $\frac{m}{2}\omega^2x^2$ in Fig.~\ref{fig:1}(b)) is tuned to the confinement potential shape (blue curve), with the corresponding period value $2\pi/\omega=5.5$~ps.
In Fig.~\ref{fig:2}(a) both spinor components are represented by two Gaussian wave packets (green $|\psi_\uparrow(x,t)|^2$ and blue $|\psi_\downarrow(x,t)|^2$ density maps) that oscillate back an forth with increasing amplitudes, and are shifted in phase by $\pi$.
Thus, we observe that CSs emerge as a result of RSOI oscillation---yellow curve in Fig.~\ref{fig:2}(b). In the simulations presented in Figs.~\ref{fig:2} and \ref{fig:3} the spin is initially aligned along the $x$ axis.


The spinor components form CSs with opposite parameters: $\alpha_{\uparrow}=-\alpha_{\downarrow}$, their modules are shown in Fig.~\ref{fig:2}(c). 
The module increase is approximately linear, as expected.
The limitation of its growth, seen for $t>10$~ps, is due to the fact that the electron enters a region where the confinement potential is no longer parabolic (see blue vs. gray curve in Fig.~\ref{fig:1}(b)).
For RSOI coupling strength $\gamma=\gamma_\mathrm{3D}|e|\langle E_y\rangle$ (yellow curve in Figs.~\ref{fig:2}(b), \ref{fig:3}(b)), with amplitude $\gamma_\mathrm{3D}|e|E_{y0}\simeq40$~meV~nm, we have $\sqrt{\frac{m\omega }{2\hbar }}\frac{\gamma_\mathrm{3D}|e|E_{y0}}{2\hbar}\simeq0.26$~ps$^{-1}$, which gives a black sloping line from Figs.~\ref{fig:2}(c) and \ref{fig:3}(c).
The CSs ``quality'' parameters $c_{\uparrow}$ and $c_{\downarrow}$ are close to unity during the simulation.
For quantum computing purposes, one needs CSs with relative high amplitudes $|\alpha|>2$,\cite{22} but using error correction techniques, the amplitudes of useful states can be reduced to $|\alpha|>1.2$.\cite{23} Here we show that we can generate states with $|\alpha|$ up to $2$. 

For a given $\alpha$, the corresponding CS expectation position is $\sqrt{2\hbar/(m\omega)}\mathrm{Re}\{\alpha\}$. 
The lines from Fig.~\ref{fig:2}(a) mark CS position calculated in three different ways: $\langle x\rangle_\uparrow=\langle \psi _{\uparrow }|x|\psi _{\uparrow }\rangle$ and $\langle x\rangle_\downarrow$ (circles), $\sqrt{2\hbar/(m\omega)}\mathrm{Re}\{\alpha_{\uparrow}\}$ and $\sqrt{2\hbar/(m\omega)}\mathrm{Re}\{\alpha_{\downarrow}\}$ (squares), $-\frac{\gamma_\mathrm{3D}|e|E_{y0}}{2\hbar} t \cos(\omega t)$ and $\frac{\gamma_\mathrm{3D}|e|E_{y0}}{2\hbar} t \cos(\omega t)$ (dashed line). The first two, based on numerical calculations, correctly track the CS position. 
The third one, analytical, is correct only up to the confinement parabolicity limit.

The oscillator frequency, seen as a frequency of the CSs position oscillation in Fig.~\ref{fig:2}(a), increases, because the curvature of the confinement parabola varies over time (changes of an induced charge distribution on the wire interfaces due to changes of the electron density\cite{47}). 
Deviations from the parabolic shape of the confinement potential, including those that appear over time, are the main constraint of the method fidelity. 
Voltages oscillation frequency $\omega/2\pi$ must correspond to the confinement harmonic oscillator frequency, while voltages oscillation amplitude $V_\mathrm{lr}$ may be arbitrarily small, much smaller than those in simulations.
However, the CS generation time is proportional to voltages amplitude $V_\mathrm{lr}$, which sets $E_{y0}$.

Both obtained CSs have a definite spin, so our scheme can be used to obtain the spin-density separation\cite{32}
by dividing the dot into two parts at the right moment.
An additional change of the voltage on the $\mathrm{e}_3$ gate depicted by a dark red curve in Fig.~\ref{fig:3}(a) (it begins to decrease from $20$~mV at $7$~ps, reaching $-60$~mV at $9.5$~ps) generates a separating potential barrier---illustrated by the green curve in Fig.~\ref{fig:1}(b).
The spin $z$-component $\frac{\hbar}{2}\langle\sigma_z\rangle$ calculated separately in the left and right half of the wire is presented in Fig.~\ref{fig:3}(b) by red and orange curves, respectively.
Hence, we can observe stabilization of these components in the later part of the simulation. This also gives stabilization of the CSs.
The choice of the moment of the lifting the separation barrier is not critical, due to the rectangular shape of the spin components course.
Now, if we measure in which part of the dot the electron is located, the result depends on the spin alignment.
This gives the opportunity to measure the electron spin.

\begin{figure}[ht!]
\includegraphics[width=8.7cm]{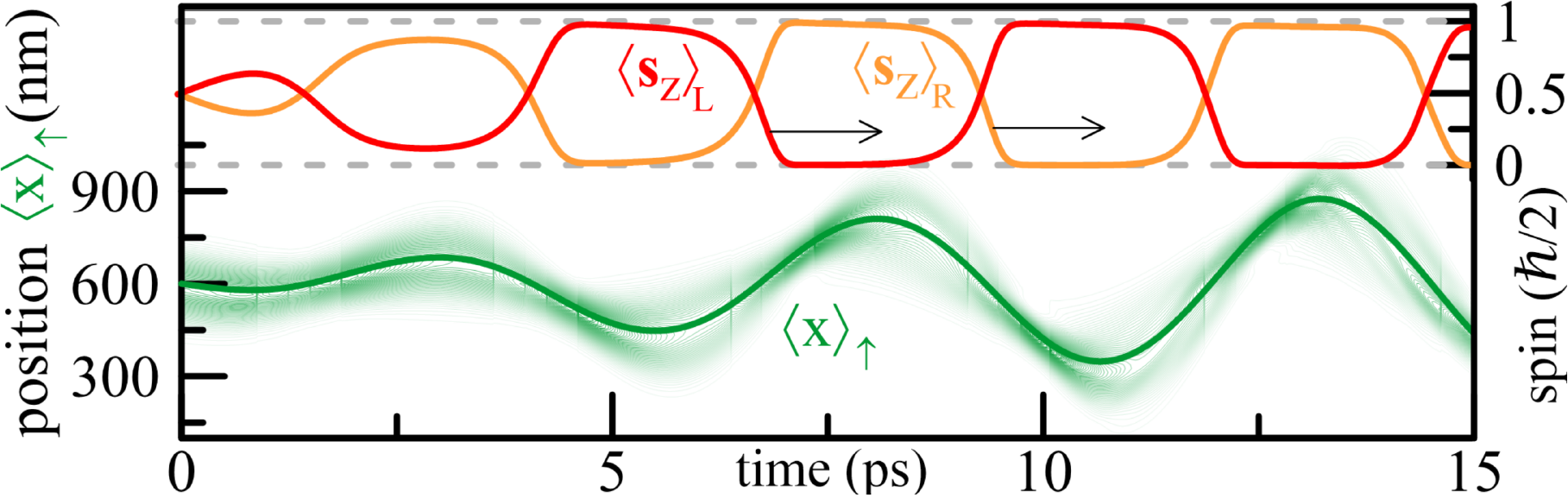}
\caption{\label{fig:4} Results for a spin initially aligned with the $z$ axis producing not a superposition but a single CS.}
\end{figure}

Figs.~\ref{fig:2} and \ref{fig:3} present simulations which lead to the Schr\"odinger cat state,
while in simulation from Fig.~\ref{fig:4} the spin is initially aligned with the $z$ axis. We observe analogous generation, but here producing a single CS.
In this way, we can initialize a coherent-state spin qubit with parameter $\alpha_\uparrow$.

\textit{Conclusions}.
We have proposed a nanodevice---a gated semiconductor nanowire---which allows for generation of spin-dependent CSs. Our proposal can be considered as a solid state spintronic analog of the famous quantum optics experiments.
The key ingredient of our proposal is a time-dependent oscillating spin-orbit interaction which induces separation of the spin-up and spin-down single electron states located on opposite sides of the wire---the so-called 'electronic' Schr\"odinger cat state. 
Effective production of such CSs by means of oscillating RSOI is done using only the electric fields generated by oscillating voltages applied to the local gates.
The proposed method is also suitable for realization of a precise single electron spin readout.


\textit{Acknowledgments}.
This work has been supported by National Science Centre, Poland, under Grant No. UMO-2014/13/B/ST3/04526. P.~S. acknowledge support from SCIEX. 



\bibliography{coh-states-gen} 

\begin{thebibliography}{80}
\bibitem [{101()}]{101}%
  \BibitemOpen
  \href@noop {} {}\bibinfo {note} {Here we assume a nanowire with zinc-blende
  crystallographic structure, grown in [111] direction where the Dresselhaus
  SOI term is expected to be negligible. The InSb Rashba SOI coefficient is
  $\gamma_\mathrm{3D}=5.23$~nm$^2$.}\BibitemShut {Stop}%
\bibitem [{103()}]{103}%
  \BibitemOpen
  \href@noop {} {}\bibinfo {note} {Detailed derivation can be found in the
  Appendix.}\BibitemShut {Stop}%
\end{thebibliography}

\appendix*

\section{1D harmonic oscillator with the present SOI in the momentum space}
The Hamiltonian (\ref{ham1d}) in momentum representation
\[H_\mathrm{1d}(p)=\frac 1{2m}\left(p^2-\left(\zeta\partial _p\right)^2\right)1_2-\frac{q\hbar}{m}p\sigma_z\]
written in the canonical form is 
\[H'_\mathrm{1d}(p')=\frac 1{2m}\left(p'^2-\left(\zeta \partial_{p'}\right)^2 1_2\right),\]
where we have skipped a constant, and $p' = p 1_2-\hbar{}q\sigma_z$, $q = m\gamma_\mathrm{3D}e E_y/\hbar^2 $, $\zeta = m\hbar \omega$.
The ground state wave function is doubly degenerate due to the spin:
\[\Psi _{\pm}(p')=\left(4\pi\zeta\right)^{-1/4}\text{exp}\left(-\frac{p'^2}{2\zeta }\right)\chi_{z\pm},\]
where $\chi_{z+}\!\!=$\footnotesize{$\left(\begin{matrix}1\\0\end{matrix}\right)$}\normalsize, $\chi_{z-}\!\!=$\footnotesize{$\left(\begin{matrix}0\\1\end{matrix}\right)$}\normalsize, 
and $E'=\hbar \omega $.
After substitution $p' = p 1_2-\hbar{}q\sigma_z$ and some matrix operations, we obtain
$\Psi_{\pm}(p)\sim\exp{\left(-\frac{p^2}{2\zeta }\right)}\exp{\left(\pm\frac{{\hbar{}qp}}{\zeta }\right)}\chi_{z\pm}$
and $E=\hbar \omega -\frac{\hbar^2q^2}{2m}$.
Finally, the inverse Fourier transform gives us the position representation ($\beta=\frac{m\omega}{2\hbar}$):
\[\Psi_{\pm}(x)=\left(\frac{\beta}{2\pi}\right)^{1/4}\!\exp{\left(-\beta x^2\right)}\exp{\left(\pm iqx \right)}\chi_{z\pm}.\]

\end{document}